# Association between nutritional factors, inflammatory biomarkers and cancer types: an analysis of NHANES data using machine learning

Yuqing Liu, M.P.H.[1], Meng Zhao, Ph.D.[2, *], Guanlan Hu, Ph.D. [3], Yuchen Zhang M.S.,[1]

[1] Department of Epidemiology, Columbia University Mailman School of Public Health, New York, NY
[2] Department of Biomedical Informatic, Columbia University, New York, NY
[3] Data Science Institute, Columbia University, New York, NY

**Abstract:**

*Background.* Diet and inflammation are increasingly recognized as critical factors influencing cancer risk, yet the combined impact of nutritional status and inflammatory biomarkers on cancer status and cancer types leveraging machine learning approaches remains insufficiently explored.

*Objectives.* This study aims to investigate association between nutritional factors, inflammatory biomarkers, and cancer status, while assessing whether these relationships differ across cancer types using data from the National Health and Nutrition Examination Survey (NHANES).

*Methods.* We analyzed 24 macro- and micronutrients, along with the inflammatory biomarker C-reactive protein (CRP) and the advanced lung cancer inflammation index (ALI) as key features in 26,409 NHANES participants (2,120 with a cancer diagnosis). To assess their influence on cancer prevalence, a multivariable logistic regression model was employed. Then, we performed analyses focusing on the five most common cancer types to determine whether the impact of these features varied significantly across different cancer types. Additionally, to evaluate the predictive value of these features for cancer status, three machine learning models: Logistic Regression, Random Forest, and XGBoost were utilized on the full feature set.

*Results.* The cohort's mean age was 49.1±18.3 years, and 34.7% classified as obese. Experimental results reveal that comorbidity such as anemia and liver conditions and nutritional factors, such as protein and several vitamins, play a crucial role in predicting cancer status. Our models achieve satisfactory performance, with Random Forest yielding the highest accuracy (0.72) among the three machine learning models.

*Conclusions.* Our findings suggest that higher-quality nutritional intake and lower inflammatory factors may have a protective effect against cancer risk. These insights could support the use of optimal therapeutic strategies for cancer prevention.

**Keywords**: nutrition, inflammation, cancer, C-reactive protein, NHANES, machine learning

# 1. Introduction

Cancer continues to be a major global health concern, with millions of new cases and deaths reported annually. Identifying individuals at higher risk of cancer early can significantly improve patient outcomes and provide more effective preventive care. While established risk factors such as age, gender, lifestyle, and comorbidities have been extensively associated with cancer incidence, more and more studies highlight the critical role of diet, nutrition, and inflammation in cancer development and progression by influencing proinflammatory carcinogenic effects or anticancer immune responses (1). Diet choices have the potential to impact the inflammatory responses and the progression of cancer. Certain food categories and dietary patterns, such as Mediterranean Diet and DASH (Dietary Approaches to Stop Hypertension) Diet, which emphasize fresh fruits, vegetables, legumes, nuts, and whole foods, have been shown to reduce inflammatory markers and pro-inflammatory conditions. Conversely, processed and refined foods have been linked to increased inflammation (2). Studies have also found that diets with relatively high saturated fatty acid, trans fatty acid, cholesterol, and vitamin $B_{12}$, and iron may be associated with elevated inflammation levels. In contrast, nutrient-rich dietary components, including flavonoids, fiber, polyunsaturated fatty acids, omega-3 and omega-6 fatty acids, turmeric, garlic, ginger, vitamins, niacin, magnesium, and zinc, have demonstrated anti-inflammatory properties (3). These findings underscore the potential role of dietary interventions in modulating inflammation and, consequently, reducing cancer risk.

The association between inflammation and tumors is also intricately connected as a growing body of data suggests that inflammation is an important factor in tumor growth (4). Prolonged inflammatory states may impede the immune system's capacity to detect and eliminate tumor cells, while simultaneously creating a milieu conducive to additional tumor cell proliferation (5). Recent studies have identified nutritional and inflammatory variables as potential predictors of a variety of cancers, including breast and liver cancers (6). Moreover, evidence shows that the dietary inflammatory index, such as C-reactive protein (CRP) in serum, has been highly associated with the risk of developing colorectal, lung, and breast cancer (1, 7). Inflammatory biomarkers, such as CRP and Advanced Lung Cancer Inflammation Index (ALI), have been associated with the risk of specific cancer types like colorectal and lung cancer (8-10). Given the intricate link between nutrition and inflammation, assessing their combined impact on cancer could provide valuable insights for developing targeted clinical interventions to reduce inflammation and lower cancer risk.

Recent advancements in statistics and machine learning techniques have significantly enhanced epidemiological research in identifying disease-associated risk factors and predicting health conditions (11, 12). Given the substantial global burden of cancer, there is a need to utilize these advanced techniques to explore the potential association between nutritional, inflammatory biomarkers, and cancer status. These factors could serve as potential early screening tools for cancer detection and reducing long-term healthcare costs. However, few studies have investigated the association between nutrition, inflammation and cancer status, and they have not accounted for cancer types, highlighting the need for further investigation. To fill up the research gap, our study is the first to assess the combined influence of nutritional and inflammatory status (specifically CRP and ALI) and cancer types based on the National Health and Nutrition Examination Survey (NHANES) (1999-2018) using machine learning methods.

## 2. Methods

*Study population*

This study utilized data extracted from the National Health and Nutrition Examination Survey (NHANES), a comprehensive program that gathers information on health and nutritional status of population in the United States through interviews, physical examinations, laboratory tests, etc (13). NHANES survey has been conducted biennially since 1999, with each cycle representing a nationally representative sample. In this study, we used publicly available data from NHANES between 1999 and 2010, including a total of 26,409 participants aged 20 and older who provided detailed information on demographics, dietary intake, clinical examinations, laboratory results, and cancer history. The dietary information was gathered from the post-dietary recall questionnaire which included all the food a participant consumed within 24 hours prior to the interview. The nutrients from the food consumed within 24 hours were summed to calculate the total nutrient content for that period. Among the participants, a subset of 2,120 individuals self-reported a cancer diagnosis. And the top five most prevalent cancer types were selected for further analysis, including skin cancer (non-melanoma) (383, 18.1%), prostate cancer (316, 14.9%), breast cancer (288, 13.6%), skin (do not know what kind) (175, 8.3%), cervix (171, 8.1%). All selected variables had less than 5% missing data, and missing values were imputed using mean imputation for numeric variables and mode imputation for categorical variables.

*Variables*

In this study, demographic variables selected in the analysis were age, gender, and race. For the nutritional variables, we used energy intake as well as the intake of key macronutrients such as protein, carbohydrate, fiber, total fat, total cholesterol, various vitamins and minerals, as well as multiple types of saturated fatty acids (SFA), monounsaturated fatty acids (MFA), and polyunsaturated fatty acids (PFA), offering a comprehensive assessment of diet and nutrition. The advanced lung cancer inflammation index (ALI) together with inflammatory biomarker C-reactive protein (CRP) were used as inflammatory variables. The ALI was calculated as $ALI = BMI \times Alb / NLR$, where BMI is the body weight in kilograms / (height in meters)$^2$ obtained from the examination data, Alb is serum albumin in grams per deciliter, and NLR is absolute neutrophil count / absolute lymphocyte count based on blood samples (8). The concentration of CRP was obtained from blood specimens in NHANES laboratory dataset and was measured using latex-enhanced nephelometry, with a normal range below 0.3 mg/dL (14).

Cancer diagnosis was self-reported and collected through the Medical Condition Question, where participants were asked "Have you ever been told by a doctor or other health professional that you have cancer or any malignancy?". Participants who answered "yes" were considered as having a history of cancer, while those who answered "no" were considered as not having experienced cancer. For patients who answered 'yes,' a follow-up question asked about the type of cancer they had.

*Statistical analysis*

First, participants in this study were divided into two groups based on their cancer status: those with cancer and those without cancer (control group), to assess the association between nutritional and inflammatory status and cancer status. Then, we selected the significant variables for a secondary analysis to assess their influence across cancer types. Categorical variables were expressed as counts and percentages, while continuous variables were summarized as the means ± standard deviations. The $t$-test and $\chi^2$ test were used to compare baseline characteristics between the two groups for continuous and categorical variables, respectively (15, 16). We use Model 1 to denote the one with no adjustment and Model 2 with adjustment for conventional variables including gender, BMI, and comorbidities such as congestive heart failure, stroke, and liver disease. We estimated the odds ratios (OR) along with P-values and 95% confidence intervals (CIs) to determine statistical significance. An outcome was considered significant if $p < 0.05$. To further evaluate the relationship between nutritional and inflammatory factors and cancer types, we selected variables with $p < 0.05$, calculated the mean concentrations of nutritional and inflammatory biomarkers within each cancer type, and conducted an ANOVA test to determine whether their levels significantly differed among the five most common cancer subtypes (skin cancer (non-melanoma); prostate cancer; breast cancer; skin cancer (unclear type); and cervix cancer) (17).

*Machine learning model*

To assess the combined predictive power of our nutritional and inflammatory features for cancer status, we trained three supervised classifiers—Logistic Regression, Random Forest, and XGBoost—using the full set of selected variables. Model performance was evaluated by calculating both the area under the receiver operating characteristic curve (AUC) and overall accuracy (18). We implemented stratified five-fold cross-validation to partition the data into training and testing subsets, ensuring balanced representation of cancer cases in each fold. Feature importance was quantified using XGBoost's built-in gain metric to highlight which predictors contributed most strongly to discrimination. All data preprocessing, statistical analyses, and machine learning workflows were executed in R (v4.3.2) within RStudio (2023.12.0) (19, 20).

## 3. Results

*3.1 Baseline characteristics*

Our finalized data comprised of 26,409 participants (age = 49.1±18.3), of which 2,120 were diagnosed with one or more types of cancer and 24,289 were without cancer. Baseline characteristics of the participants were presented by cancer status in Table 1. Compared with the control group, people who experienced cancer were more likely to be older, White, experiencing congestive heart failure, liver disease and stoke ($p < 0.001$). In contrast, there were no significant statistical differences in gender and BMI between the two groups. Additionally, there were significant differences in the inflammatory biomarker CRP and ALI between the cancer and control groups ($p < 0.001$). Among the nutritional indicators, energy, protein, carbohydrate, total fat, cholesterol, niacin, calcium, phosphorus, magnesium, zinc, Copper, sodium, and selenium showed significant differences between the participants who were experiencing cancer and those who were not ($p < 0.001$).

*3.2 Association between nutrition, inflammation biomarkers, and cancer*

The association between nutrition, inflammation biomarkers, and cancer status was analyzed by the multivariate logistic regression model and the results were presented in **Table 2**. The forest plots (**Figure 1a, Figure 1b**) illustrated these associations for better visualization. In both unadjusted (Model 1) and adjusted (Model 2) models, several nutritional factors demonstrated statistically significant associations with cancer. Notably, higher energy intake was consistently linked to a lower likelihood of cancer (OR = 0.99973, 95% CI: 0.99968–0.99979, $p$ <2e-16). Similarly, protein, carbohydrate, total fat, and cholesterol intake were all inversely associated with cancer, suggesting potential protective effects. Among vitamins and minerals, niacin (OR = 0.989, 95% CI: 0.985–0.993, $p$ <2.8e-08), phosphorus (OR = 0.9998, 95% CI: 0.9997–0.9999, $p$ < 2.87e-08), and selenium (OR = 0.997, 95% CI: 0.996–0.998, $p$ < 4.60e-11) also exhibited significant negative associations with cancer prevalence. Vitamin C had protective effect for cancer (OR = 0.9994, 95% CI: 0.9989–0.9999, $p$ = 0.01466). Conversely, vitamin A showed a slight positive association with cancer risk in both models, although its effect size was minimal (OR = 1.000051, 95% CI: 1.000008–1.000090, $p$ = 0.013751). Minerals including Calcium, Phosphorus, Magnesium, Zinc, Copper, Sodium, Selenium were negatively associated with cancer with ORs of 0.99989, 0.9998, 0.9997, 0.9919, 0.948, 0.99990, and 0.997 respectively. Inflammatory biomarkers also played a notable role in cancer risk. C-reactive protein (CRP) was significantly associated with increased cancer risk in both models (OR = 1.06, 95% CI: 1.01–1.10, $p$ = 0.0047), reinforcing the well-established link between systemic inflammation and cancer progression. Conversely, the Advanced Lung Index (ALI) demonstrated an inverse association with cancer (OR = 0.997, 95% CI: 0.996–0.999, $p$ = 8.04e-05), suggesting that a higher ALI may be indicative of a lower cancer risk.

When examining differences in nutrition, dietary intake, and inflammatory biomarker levels among five different cancer subtypes—skin cancer (non-melanoma), prostate cancer, breast cancer, skin cancer (unclear type), and cervix cancer (**Table 3**)—we found that all nutritional levels, such as energy, protein, carbohydrate concentrations, vitamins, and minerals, differed significantly among the cancer subtypes ($p < 0.001$). Each cancer subtype exhibited distinct patterns of nutrient intake. However, dietary and inflammatory biomarkers, including CRP and ALI, remained relatively consistent across the five cancer subtypes, with only minor variations among the subgroups.

*3.3 Performance of Machine Learning Classification*

Machine learning is an advanced form of pattern recognition that enables machines to make judgments by analyzing large amounts of data. In this work, we tested three machine learning methods, and the results highlighted comorbidities such as liver conditions, and CRP as significant predictors of cancer. Nutrition variables like energy and protein intake also played a role in prediction. XGBoost yielded similar results, with protein intake emerging as the most important features. XGBoost also identified interactions between nutrition and comorbidities. A comparison of model performance (**Table 5**) demonstrates that Random Forest outperformed both Logistic Regression and XGBoost across multiple metrics, achieving the highest accuracy (0.72), along with strong precision (0.70), recall (0.72), and F1-score (0.71).

## Discussion

In this nationally representative NHANES cohort, we demonstrated that both nutritional intake and systemic inflammation are independently and jointly associated with cancer prevalence. Multivariable analyses revealed that higher energy, protein, and several micronutrient intakes were inversely associated with cancer status, while elevated C-reactive protein (CRP) levels conferred increased odds. The Advanced Lung Cancer Inflammation Index (ALI) likewise showed a protective inverse relationship. Machine learning classifiers—particularly Random Forest—successfully integrated these diverse data streams to achieve robust discrimination (accuracy = 0.72, F1-score = 0.71), underscoring the complementary roles of diet and inflammation in cancer risk stratification. Our application of machine learning to NHANES data illustrates the feasibility of deploying modern predictive algorithms on large, complex epidemiologic datasets. By implementing stratified five-fold cross-validation and rigorous hyperparameter tuning within each fold, we ensured model generalizability and minimized overfitting. The superior performance of tree-based ensembles highlights their capacity to capture nonlinear relationships and interactions inherent in high-dimensional public-health surveys.

The results suggest that a diet higher in energy, protein, niacin, Vitamin C and minerals may help reduce cancer risk. Additionally, CRP and ALI could serve as useful predictive factors for cancer with lower variations among cancer subtypes. CRP was found to be positively associated with cancer, consistent with previous findings (21-23). ALI demonstrated a significant protective effect against cancer. Higher ALI leveld were associated with lower risk of cancer. The consistency of CRP and ALI across the five cancer subtypes indicates that their levels are similar in each subtype, suggesting their potential as generic predictors for cancer. Future analyses could explore the reasons behind variations in nutritional concentrations among cancer subtypes, as well as the biological mechanisms through which these nutrients influence cancer development. Additionally, NHANES did not record some classic inflammatory factors such as interleukin-6, interleukin-10, etc. Thus, future analysis could also incorporate these inflammatory factors. This study utilized cross-sectional data from NHANES, which lacks a temporal component, limiting the ability to assess causality. For example, the study found a slightly negative association between cancer and sodium intake. It is possible that cancer patients have modified their diets to reduce sodium consumption after being diagnosed. Therefore, a temporal analysis of patients' diets and cancer progression is needed. Future research should consider using longitudinal datasets that account for temporal relationships to better understand changes over time.

Several limitations warrant consideration. The cross-sectional design precludes causal inference, and cancer status was self-reported, raising the possibility of misclassification. Dietary intake was captured via a single 24-hour recall, which may not reflect long-term habits. Although missing data were minimal and imputed conservatively, residual bias cannot be excluded. Future work should validate these findings in prospective cohorts with repeated dietary assessments and clinical cancer diagnoses. Incorporating additional inflammation markers (e.g., interleukins) and leveraging deep-learning architectures may further refine risk prediction. Ultimately, our integrative, machine-learning–driven approach lays a foundation for personalized nutritional and anti-inflammatory strategies in cancer prevention.

# References

1. Zitvogel L, Pietrocola F, Kroemer G. Nutrition, inflammation and cancer. Nature immunology. 2017; 18:843-50.
2. Tristan Asensi M, Napoletano A, Sofi F, Dinu M. Low-grade inflammation and ultra-processed foods consumption: a review. Nutrients. 2023; 15:1546.
3. Stumpf F, Keller B, Gressies C, Schuetz P. Inflammation and nutrition: friend or foe? Nutrients. 2023; 15:1159.
4. Chai EZP, Siveen KS, Shanmugam MK, Arfuso F, Sethi G. Analysis of the intricate relationship between chronic inflammation and cancer. Biochemical Journal. 2015; 468:1-15.
5. Elinav E, Nowarski R, Thaiss CA, Hu B, Jin C, Flavell RA. Inflammation-induced cancer: crosstalk between tumours, immune cells and microorganisms. Nature Reviews Cancer. 2013; 13:759-71.
6. Aleksandrova K, Boeing H, Nöthlings U, Jenab M, Fedirko V, Kaaks R, et al. Inflammatory and metabolic biomarkers and risk of liver and biliary tract cancer. Hepatology. 2014; 60:858-71.
7. Nanri A, Moore MA, Kono S. Impact of C-reactive protein on disease risk and its relation to dietary factors: literature review. Asian Pacific Journal of Cancer Prevention. 2007; 8:167.
8. Jafri SH, Shi R, Mills G. Advance lung cancer inflammation index (ALI) at diagnosis is a prognostic marker in patients with metastatic non-small cell lung cancer (NSCLC): a retrospective review. BMC cancer. 2013; 13:1-7.
9. Tomita M, Ayabe T, Maeda R, Nakamura K. Combination of advanced lung cancer inflammation index and C-reactive protein is a prognostic factor in patients with operable non-small cell lung cancer. World Journal of Oncology. 2017; 8:175.
10. Pang H-Y, Chen X-F, Yan M-H, Chen L-H, Chen Z-X, Zhang S-R, et al. Clinical significance of the advanced lung cancer inflammation index in gastrointestinal cancer patients: a systematic review and meta-analysis. Frontiers in Oncology. 2023; 13:1021672.
11. Ming C, Viassolo V, Probst-Hensch N, Chappuis PO, Dinov ID, Katapodi MC. Machine learning techniques for personalized breast cancer risk prediction: comparison with the BCRAT and BOADICEA models. Breast Cancer Research. 2019; 21:1-11.
12. Yaqoob A, Musheer Aziz R, verma NK. Applications and techniques of machine learning in cancer classification: a systematic review. Human-Centric Intelligent Systems. 2023; 3:588-615.
13. Fain JA. NHANES: use of a free public data set. SAGE Publications Sage CA: Los Angeles, CA; 2017. p. 151-.
14. Nehring SM, Goyal A, Patel BC. C reactive protein. 2017.
15. Kim TK. T test as a parametric statistic. Korean journal of anesthesiology. 2015; 68:540-6.
16. Tallarida RJ, Murray RB, Tallarida RJ, Murray RB. Chi-square test. Manual of pharmacologic calculations: with computer programs. 1987:140-2.
17. Girden ER. ANOVA: Repeated measures: sage; 1992.
18. Huang J, Ling CX. Using AUC and accuracy in evaluating learning algorithms. IEEE Transactions on knowledge and Data Engineering. 2005; 17:299-310.
19. R Core Team R. R: A language and environment for statistical computing. 2013.
20. Wickham H, François R, Henry L, Müller K. dplyr: A Grammar of Data Manipulation. R package version 0.7. 6. 2018.

21. Allin KH, Nordestgaard BG. Elevated C-reactive protein in the diagnosis, prognosis, and cause of cancer. Critical reviews in clinical laboratory sciences. 2011; 48:155-70.
22. Allin KH, Bojesen SE, Nordestgaard BG. Baseline C-reactive protein is associated with incident cancer and survival in patients with cancer. Journal of clinical oncology. 2009; 27:2217-
23. Mahmoud FA, Rivera NI. The role of C-reactive protein as a prognostic indicator in advanced cancer. Current oncology reports. 2002; 4:250-5.

# Figure and Tables

**Table 1. Baseline characteristics of study population by cancer status**

| Variables | Overall (26,409) | Control (24,289) | Cancer (2,120) | p-value |
|---|---|---|---|---|
| **Age, years** | 49.1 (±18.3) | 47.8 (±17.9) | 65.1 (±15.1) | < 0.001 |
| **Male, n (%)** | 12,712 (48.1%) | 11,715 (48.2%) | 997 (47.0%) | 0.287 |
| **Race, n (%)** | | | | < 0.001 |
| Mexican American | 5,501 (20.8%) | 5,343 (22.0%) | 158 (7.5%) | |
| Other Hispanic | 1,774 (6.7%) | 1,707 (7.0%) | 67 (3.2%) | |
| White | 13,142 (49.8%) | 11,538 (47.5%) | 1,604 (75.6%) | |
| Black | 4,957 (18.8%) | 4,706 (19.4%) | 251 (11.8%) | |
| Other | 1,035 (3.9%) | 995 (4.1%) | 40 (1.9%) | |
| **BMI, $kg/m^2$** | 28.7 (±6.5) | 28.7 (±6.5) | 28.5 (±6.2) | 0.504 |
| **BMI, n (%)** | | | | 0.576 |
| Healthy | 7,934 (30.0%) | 7,294 (30.0%) | 640 (30.2%) | |
| Overweight | 9,311 (35.3%) | 8,537 (35.2%) | 774 (36.5%) | |
| Obese | 9,164 (34.7%) | 8,458 (34.8%) | 706 (33.3%) | |
| **Comorbidities** | | | | |
| CHF | 785 (3.0%) | 629 (2.6%) | 156 (7.4%) | < 0.001 |
| Liver Disease | 841 (3.2%) | 753 (3.1%) | 88 (4.1%) | 0.008 |
| Stroke | 875 (3.3%) | 715 (3.0%) | 160 (7.6%) | < 0.001 |
| **Nutrition** | | | | |
| Energy | 2116.4 (±1012.3) | 2136.0 (±1023.6) | 1891.4 (±840.4) | < 0.001 |
| Protein | 80.5 (±42.5) | 81.27 (±43.1) | 71.9 (±34.3) | < 0.001 |
| Carbohydrate | 260.7 (±130.4) | 263.0 (±131.8) | 234.2 (±110.0) | < 0.001 |
| Fiber | 16.1 (±10.0) | 16.1 (±10.1) | 16.0 (±9.4) | 0.898 |
| Total fat | 78.8 (±46.1) | 79.4 (±46.6) | 71.8 (±39.6) | < 0.001 |
| Cholesterol | 292.9 (±241.7) | 295.7 (±243.6) | 261.4 (±216.2) | < 0.001 |
| Vitamin $E$ | 7.3 (±5.7) | 7.3 (±5.7) | 7.3 (±6.1) | 0.986 |
| Vitamin $A$ | 655.8 (±871.3) | 652.1 (±880.6) | 698.5 (±754.9) | 0.019 |
| Vitamin $B_1$ | 1.6 (±0.9) | 1.6 (±0.9) | 1.6 (±0.9) | 0.022 |
| Vitamin $B_2$ | 2.1 (±1.2) | 2.1 (±1.3) | 2.1 (±1.1) | 0.793 |
| Niacin | 23.7 (±14.0) | 23.9 (±14.2) | 21.8 (±11.5) | < 0.001 |
| Vitamin $B_6$ | 1.9 (±1.3) | 1.9 (±1.3) | 1.9 (±1.2) | 0.014 |
| Folate | 505.5 (±349.1) | 506.2 (±350.2) | 497.2 (±335.8) | 0.257 |
| Vitamin $B_{12}$ | 5.2 (±7.9) | 5.2 (±8.0) | 5.2 (±6.7) | 0.887 |
| Vitamin $C$ | 93.3 (±105.3) | 93.8 (±106.1) | 87.1 (±95.4) | 0.005 |
| Calcium | 876.0 (±590.0) | 879.8 (±106.1) | 831.8 (±523.1) | < 0.001 |
| Phosphorus | 1305.8 (±667.4) | 1314.1 (±674.2) | 1210.7 (±576.0) | < 0.001 |
| Magnesium | 284.0 (±146.1) | 284.9 (±595.4) | 274.1 (±130.7) | < 0.001 |
| Iron | 15.2 (±8.9) | 15.2 (±8.9) | 15.0 (±8.8) | 0.360 |
| Zinc | 11.7 (±9.9) | 11.8 (±10.1) | 11.0 (±7.2) | < 0.001 |
| Copper | 1.3 (±1.2) | 1.3 (±1.2) | 1.2 (±0.9) | 0.012 |
| Sodium | 3351.2 (±1819.1) | 3376.7 (±1839.3) | 3059.2 (±1538.9) | < 0.001 |
| Potassium | 2660.1 (±1302.3) | 2664.0 (±1313.1) | 2616.1 (±1171.4) | 0.105 |
| Selenium | 108.1 (±63.5) | 108.8 (±64.0) | 97.8 (±56.0) | < 0.001 |
| **CRP** | 0.46 (±0.9) | 0.45 (±0.8) | 0.53 (±1.0) | < 0.001 |
| **ALI** | 67.7 (±45.1) | 68.1 (±42.1%) | 63.6 (±70.3) | < 0.001 |

*Abbreviations*: BMI, body mass index; CHF, congestive heart failure; CRP, C-reactive protein; ALI, advanced lung cancer inflammation index.

**Table 2. Association between dietary and inflammatory variables and cancer status**

|  | Model 1 | | Model 2 | |
| --- | --- | --- | --- | --- |
|  | OR (95% CI) | p-value | OR (95% CI) | p-value |
| **Nutrition** | | | | |
| Energy | **0.99971 (0.99966, 0.99976)** | <2e-16 | **0.99973 (0.99968, 0.99979)** | < 2e-16 |
| Protein | **0.994 (0.992, 0.995)** | <2e-16 | **0.9944 (0.9931, 0.9957)** | < 2e-16 |
| Carbohydrate | **0.9980 (0.9976, 0.9984)** | <2e-16 | **0.9982 (0.9978, 0.9986)** | < 2e-16 |
| Fiber | 0.9997 (0.9952, 1.0041) | 0.898 | 1.001 (0.997, 1.006) | 0.425 |
| Total fat | **0.995 (0.995, 0.997)** | 2.89e-13 | **0.996 (0.995, 0.997)** | 9.63e-10 |
| Cholesterol | **0.9994 (0.9991, 0.9995)** | 3.61e-10 | **0.9993 (0.9991, 0.9996)** | 1.69e-08 |
| Vitamin $E$ | 1.00007 (0.99216, 1.00767) | 0.986 | 1.0037 (0.9959, 1.0113) | 0.338 |
| Vitamin $A$ | **1.000048 (1.000005, 1.000087)** | 0.0203 | **1.000051 (1.000008, 1.000090)** | 0.013751 |
| Vitamin $B_1$ | **0.943 (0.896, 0.991)** | 0.0214 | 0.970 (0.921, 1.020) | 0.23629 |
| Vitamin $B_2$ | 0.995 (0.959, 1.031) | 0.793 | 1.0150 (0.9778, 1.0521) | 0.42663 |
| Niacin | **0.987 (0.984, 0.991)** | 2.45e-11 | **0.989 (0.985, 0.993)** | 2.8e-08 |
| Vitamin $B_6$ | **0.954 (0.919, 0.989)** | 0.0139 | 0.9734 (0.9364, 1.0101) | 0.16363 |
| Folate | 0.9999 (0.9998, 1.0001) | 0.257 | 1.0000 (0.9998, 1.0001) | 0.9552 |
| Vitamin $B_{12}$ | 0.9995 (0.9932, 1.0048) | 0.888 | 1.0009 (0.9949, 1.0059) | 0.727 |
| Vitamin $C$ | **0.9993 (0.9987, 0.9998)** | 0.00477 | **0.9994 (0.9989, 0.9999)** | 0.01466 |
| Calcium | **0.99985 (0.99977, 0.99993)** | 0.000319 | **0.99989 (0.99981, 0.99997)** | 0.00958 |
| Phosphorus | **0.99974 (0.99967, 0.99981)** | 7.55e-12 | **0.9998 (0.9997, 0.9999)** | 2.87e-08 |
| Magnesium | **0.9995 (0.9991, 0.9998)** | 0.00114 | 0.9997 (0.9993, 0.9999) | 0.05043 |
| Iron | 0.998 (0.993, 1.003) | 0.360 | 1.0003 (0.9951, 1.0054) | 0.9000 |
| Zinc | **0.988 (0.983, 0.995)** | 0.000453 | **0.9919 (0.9856, 0.9978)** | 0.0106 |
| Copper | **0.925 (0.871, 0.977)** | 0.00818 | **0.948 (0.896, 0.996)** | 0.054 |
| Sodium | **0.99989 (0.99986, 0.99992)** | 1.22e-14 | **0.99990 (0.99987, 0.99993)** | 6.69e-11 |
| Potassium | 1.00003 (0.99999, 1.00006) | 0.104 | 1.00001 (0.99997, 1.00005) | 0.62995 |
| Selenium | **0.996 (0.996, 0.998)** | 1.1e-14 | **0.997 (0.996, 0.998)** | 4.60e-11 |
| **Inflammatory** | **Biomarker** | | | |
| CRP | **1.08 (1.03, 1.12)** | 0.000233 | **1.06 (1.01, 1.10)** | 0.0047 |
| ALI | **0.997 (0.995, 0.998)** | 1.41e-06 | **0.997 (0.996, 0.999)** | 8.04e-05 |

Model 1: multiple logistic regression prediction model without adjustment for covariate.
Model 2: multiple logistic regression prediction model after adjusting covariates (including gender, BMI category, congestive heart failure, stroke, liver condition)
Bold indicates significant odds ratio.
The sample size of this data is N = 26,409.

**Table 3. Cancer subtypes and dietary and inflammatory variables.**

|  | Skin cancer (non-melanoma) (383, 18.1%) | Prostate cancer (316, 14.9%) | Breast cancer (288, 13.6%) | Skin cancer (unclear what kind) (175, 8.3%) | Cervix cancer (171, 8.1%) | *p*-value |
|---|---|---|---|---|---|---|
| **Nutrition** | | | | | | |
| Energy | 1975.2 | 1978.0 | 1627.7 | 1989.7 | 1891.8 | **< 0.001** |
| Protein | 77.0 | 76.6 | 62.6 | 74.7 | 66.4 | **< 0.001** |
| Carbohydrate | 242.6 | 240.4 | 209.2 | 242.7 | 237.1 | **< 0.001** |
| Total fat | 74.5 | 74.5 | 59.5 | 79.1 | 71.2 | **< 0.001** |
| Cholesterol | 255.3 | 298.9 | 208.5 | 275.8 | 266.6 | **< 0.001** |
| Vitamin $A$ | 740.2 | 804.3 | 662.5 | 686.8 | 503.6 | **< 0.001** |
| Vitamin $B_1$ | 1.73 | 1.64 | 1.35 | 1.65 | 1.31 | **< 0.001** |
| Niacin | 23.8 | 23.1 | 18.6 | 22.6 | 18.9 | **< 0.001** |
| Vitamin $B_6$ | 2.11 | 1.99 | 1.60 | 1.99 | 1.50 | **< 0.001** |
| Vitamin $C$ | 101.6 | 93.1 | 85.0 | 83.3 | 72.8 | **0.015** |
| Calcium | 905.8 | 799.1 | 763.6 | 903.4 | 767.7 | **< 0.001** |
| Phosphorus | 1310.1 | 1237.9 | 1058.0 | 1283.3 | 1104.1 | **< 0.001** |
| Magnesium | 305.2 | 282.5 | 247.0 | 281.9 | 242.5 | **< 0.001** |
| Zinc | 12.0 | 11.7 | 9.8 | 11.3 | 9.9 | **< 0.001** |
| Copper | 1.34 | 1.29 | 1.11 | 1.24 | 1.11 | **< 0.001** |
| Sodium | 3265.2 | 3208.7 | 2688.7 | 3265.5 | 2897.2 | **< 0.001** |
| Selenium | 103.1 | 104.6 | 85.8 | 97.6 | 89.3 | **< 0.001** |
| **Dietary and Inflammatory Biomarker** | | | | | | |
| CRP | 0.39 | 0.55 | 0.47 | 0.45 | 0.55 | 0.121 |
| ALI | 59.5 | 62.0 | 58.3 | 57.6 | 67.5 | 0.423 |

**Table 4. Correlation analysis between factors and cancer subtypes**

|  | Skin cancer (non-melanoma) (383, 18.1%) | Prostate cancer (316, 14.9%) | Breast cancer (288, 13.6%) | Skin cancer (unclear what kind) (175, 8.3%) | Cervix cancer (171, 8.1%) | p-value |
|---|---|---|---|---|---|---|
| **Nutrition** | | | | | | |
| Energy | 1975.2 (±836.1) | 1978 (±870.8) | 1627.7 (±594.3) | 1989.7 (±862.2) | 1891.8 (±808.6) | < 0.001 |
| Protein | 77 (±34.7) | 76.6 (±33.2) | 62.6 (±26.5) | 74.7 (±34.6) | 66.4 (±31.7) | < 0.001 |
| Carbohydrate | 242.6 (±120.1) | 240.4 (±104.5) | 209.2 (±85.9) | 242.7 (±103.6) | 237.1 (±117.4) | 0.0001 |
| Total fat | 74.5 (±37.2) | 74.5 (±37.4) | 59.5 (±26.7) | 79.1 (±43.3) | 71.2 (±38.2) | < 0.001 |
| Cholesterol | 255.3 (±225.2) | 298.9 (±233.3) | 208.5 (±151.9) | 275.8 (±229.9) | 266.6 (±208.9) | < 0.001 |
| Vitamin $A$ | 740.2 (±575.2) | 804.3 (±1016) | 662.5 (±799.5) | 686.8 (±446.3) | 503.6 (±436.8) | < 0.001 |
| Vitamin $B_1$ | 1.7 (±1.0) | 1.6 (±0.8) | 1.3 (±0.6) | 1.7 (±0.8) | 1.3 (±0.7) | < 0.001 |
| Niacin | 23.8 (±12.2) | 23.1 (±11.4) | 18.6 (±8.5) | 22.6 (±11.6) | 18.9 (±9.3) | < 0.001 |
| Vitamin $B_6$ | 2.1 (±1.5) | 2 (±1.1) | 1.6 (±0.9) | 2 (±1.4) | 1.5 (±1.0) | < 0.001 |
| Vitamin $C$ | 101.6 (±136.4) | 93.1 (±77.8) | 85 (±79.3) | 83.3 (±76.6) | 72.8 (±91.3) | < 0.001 |
| Calcium | 905.8 (±575.6) | 799.1 (±439.8) | 763.6 (±409.5) | 903.4 (±568.5) | 767.7 (±513.9) | < 0.001 |
| Phosphorus | 1310.1 (±614.4) | 1237.9 (±514.6) | 1058 (±421.7) | 1283.3 (±621.4) | 1104.1 (±552.9) | < 0.001 |
| Magnesium | 305.2 (±140.1) | 282.5 (±124.0) | 246.9 (±102.3) | 281.9 (±144) | 242.5 (±115.9) | < 0.001 |
| Zinc | 12 (±6.9) | 11.7 (±6.6) | 9.8 (±6.1) | 11.3 (±6.5) | 9.9 (±7.7) | < 0.001 |
| Copper | 1.3 (±0.7) | 1.3 (±0.7) | 1.1 (±0.6) | 1.2 (±0.7) | 1.1 (±0.7) | < 0.001 |
| Sodium | 3265.2 (±1663.2) | 3208.7 (±1542.3) | 2688.7 (±1259.3) | 3265.5 (±1709.4) | 2897.2 (±1432.6) | < 0.001 |
| Selenium | 103.1 (±53.2) | 104.6 (±55.8) | 85.8 (±42.6) | 97.6 (±45.1) | 89.3 (±47.2) | < 0.001 |
| **Inflammation** | | | | | | |
| CRP | 0.4 (±0.8) | 0.5 (±1.4) | 0.5 (±0.7) | 0.5 (±0.8) | 0.5 (±0.7) | 0.0001 |
| ALI | 59.5 (±29.9) | 62 (±103.8) | 58.3 (±33.7) | 57.6 (±34.8) | 67.5 (±35.2) | < 0.001 |

**Table 5. Machine learning performance of predicting cancer statues**

| Model | Accuracy | Precision | Recall | F1-Score |
|---|---|---|---|---|
| Logistic Regression | 0.61 | 0.58 | 0.60 | 0.62 |
| Random Forest | 0.72 | 0.70 | 0.72 | 0.71 |
| XGBoost | 0.65 | 0.64 | 0.68 | 0.68 |

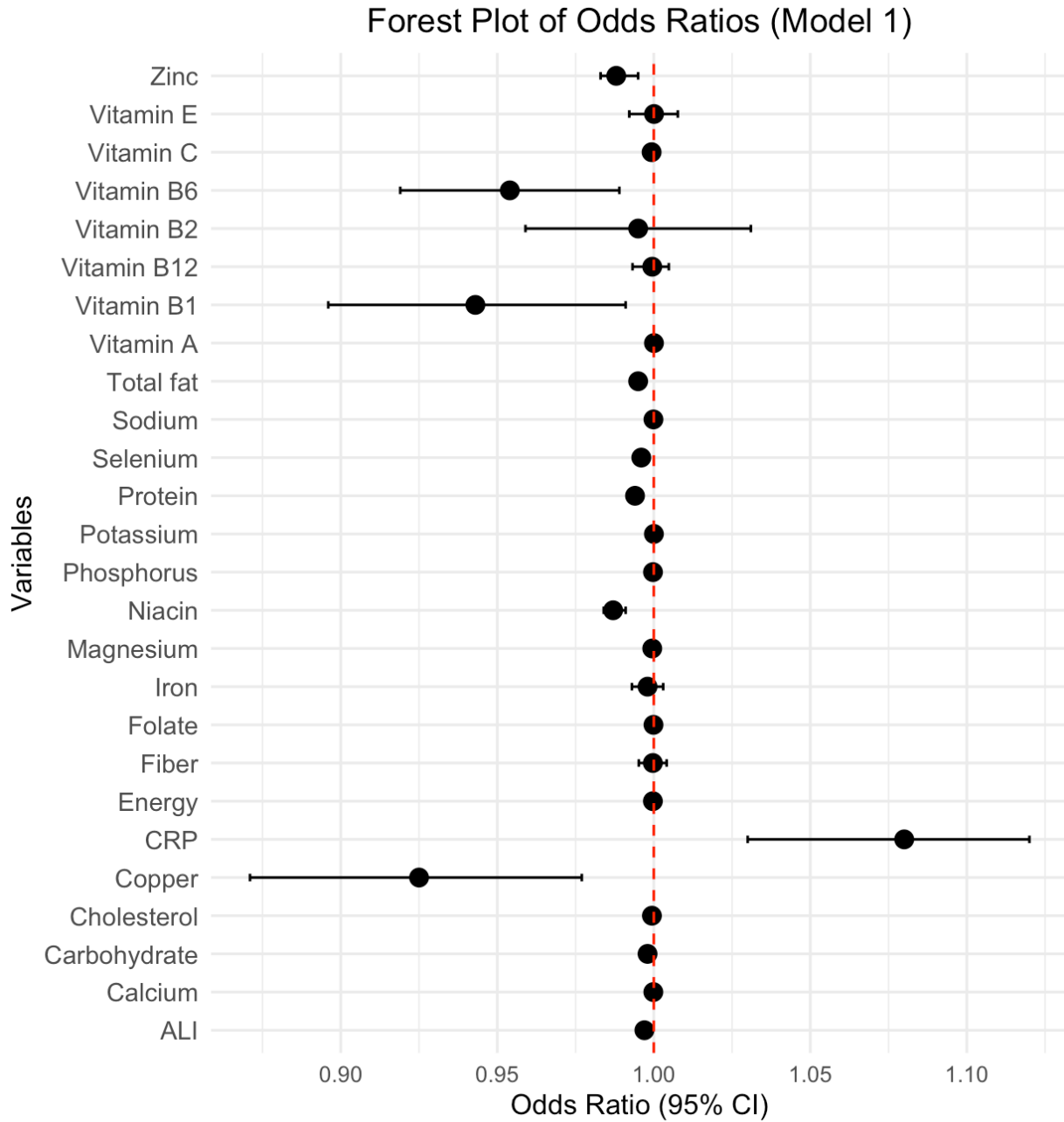

**Figure 1a. Forest plot of logistic regression results showing the association between nutrients and immflamtory biomarkers and cancer, not adjusted for covariates.**

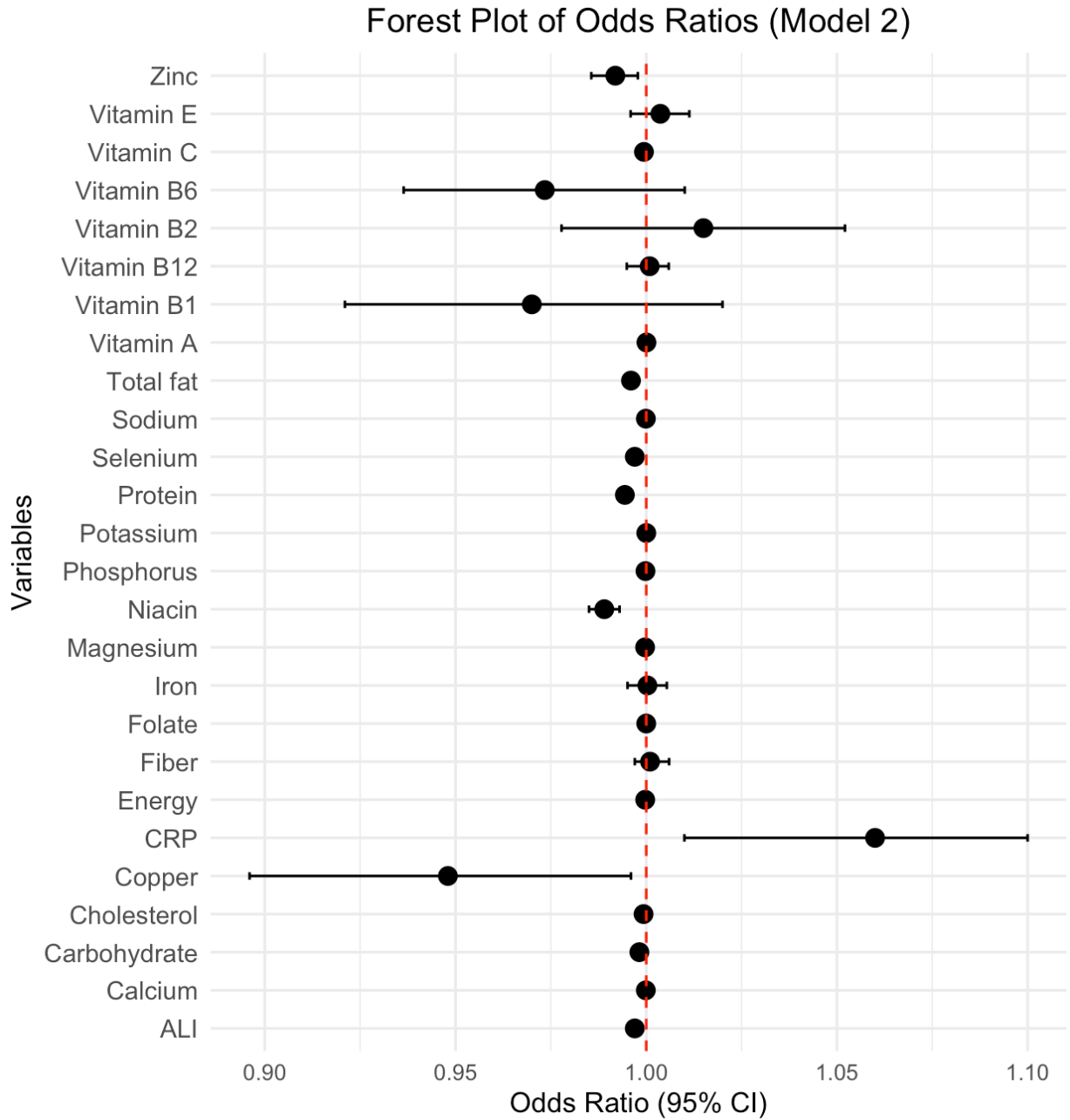

**Figure 1b. Forest plot of logistic regression results showing the association between nutrients and immflamtory biomarkers and cancer, adjusted for covariates.**